\documentclass[letterpaper,11pt]{article}

\font\titlefont=cmbx10 scaled \magstep2
\usepackage{graphics}

\begin{document}
\input{epsf}

\begin{flushright}
\vspace*{-2cm}
gr-qc/9905012 \\ 
May 4, 1999
\vspace*{1cm}
\end{flushright}

\begin{center}
{\titlefont Fluctuations of the Hawking Flux}
\vskip .4in
C.-H. Wu\footnote{email: wu@cosmos2.phy.tufts.edu} and 
L.H. Ford\footnote{email: ford@cosmos2.phy.tufts.edu} 
\vskip .2in
Institute of Cosmology\\
Department of Physics and Astronomy\\
Tufts University\\
Medford, Massachusetts 02155
\end{center}

\vskip .3in


\begin{abstract}
The fluctuations of the flux radiated by an evaporating black hole
will be discussed. Two approaches to this problem will be adopted.
In the first, the squared flux operator is defined by normal ordering.
In this case, both the mean flux and the mean squared flux are well
defined local quantites. It is shown that the flux undergoes large 
fluctuations on a time scale of the order of the black hole's mass.
Thus the semiclassical theory  of gravity, in which a classical gravitational
field is coupled to the expectation value of the stress tensor, breaks
down below this time scale. In the second approach, one does not attempt
to give meaning to the squared flux as a local quantity, but only as a 
time-averaged quantity. In both approaches, the mean squared mass
minus the square of the mean mass grows linearly in time, but four
times as fast in the second approach as in the first.
\end{abstract}

\section{Introduction}
\label{sec:Intro}

One of the most remarkable theoretical discoveries of recent decades
was that of black hole evaporation by Hawking \cite{Hawking74,Hawking} in 1974.
This discovery demonstrated that a black hole emits a (filtered)
thermal spectrum of radiation, and will eventually cease to exist as
a classical black hole. However, Hawking's derivation of this effect and
most of the subsequent  papers on the topic have dealt only with the
mean flux of radiation emitted by the black hole. There should in 
addition be fluctuations of the flux, which will be the topic of this paper.

In order to discuss the fluctuations in the components of the stress
tensor, it is necessary to be able to define the expectation value of
the square of a stress tensor component. As will be discussed in the 
following section, this is a more difficult problem than the definition
of the expectation value of the stress tensor operator itself. In flat
spacetime, one possibility is to normal order the product of stress tensor
operators. This gives a meaning to quantities such as the square of the
energy density or pressure at a spacetime point and hence to a local measure
of the fluctuations in these quantites. It will be shown in Sect.~\ref{sec:NO}
that normal ordering here involves dropping both a divergent vacuum term 
and a divergent, state-dependent, cross term.
The normal ordering  approach was adopted in
Refs. \cite{F82,DF88,Kuo}, and was generalized to static curved 
spacetimes by Phillips and Hu \cite{PH97}. It leads to the correct classical
limit in that the expectation value of a squared stress tensor component
is the square of its expectation value in the limit that the quantum state
is a coherent state. Furthermore, it leads to the prediction that quantum
states which exhibit negative energy densities have large energy
density fluctuations \cite{Kuo}.

Black hole evaporation necessarily involves negative energy density in 
that a flux of negative energy going down the horizon is needed to account
for the mass loss by the black hole. Thus the results of Ref. \cite{Kuo}
lead us to suspect that there must be large instantaneous fluctuations
in the Hawking flux. Such large fluctuations can also be inferred on
statistical physics grounds, as will be discussed in 
Sect. \ref{sec:thermalfluct}.
In the present paper, we will restrict our attention to the Hawking
flux in the asymptotic region far from the event horizon where spacetime
is approximately flat. In Sect.~\ref{sec:FFno}, this will be done using the 
normal ordering prescription. In Sect.~\ref{sec:2dmirror},
 a formula will be derived for the
squared flux radiated by a moving mirror in two dimensional spacetime.
This can be used to discuss the fluctuations in the flux enitted by a
black hole in two dimensions. In Sect.~\ref{sec:4dBH}, the corresponding 
analysis will be carried out for a Schwarzschild black hole in four dimensional 
spacetime. In both cases it will be shown that there are fractional flux 
fluctuations of order unity on time scales of the order of the mass of the 
black hole.

However, there are alternatives to the normal ordering method which
involve space or time averages of the stress tensor. One such alternative 
was used by Barton \cite{Barton} to study the fluctuations of the Casimir
force. In Sect.~\ref{sec:cross} the possibility of averaging the flux over 
time in a two dimensional model will be examined. Using an integration by parts 
procedure, it is possible to define time integrals of the state-dependent
cross term. It will be shown that 
the mean squared mass of the black hole undergoes a random walk in that
its deviation from the square of the mean mass grows linearly in time.
This is true regardless of whether or not the cross term is retained,
but the rate of growth due to the cross term is three times that
obtained in the normal ordering prescription.

The results of the paper will be summarized and discussed in 
Sect.~\ref{sec:final}.

\section{Normal-ordering expansions and the cross term}
\label{sec:NO}

In Minkowski spacetime we renormalize the expectation value of the
energy-momentum tensor by subtracting out the Minkowski divergence, 
\begin{eqnarray}
\langle T_{\mu \nu }\rangle _{renormalized} & = & \langle :T_{\mu \nu 
}:\rangle \nonumber \\
 & = & \langle T_{\mu \nu }\rangle -\langle T_{\mu \nu }\rangle_{M}\, .
\end{eqnarray}
Here $\langle \rangle_{M}$ denotes the expectation value in the Minkowski
vacuum state.
For quadratic operators, this subtraction is just normal ordering.
However, this is not true for the squared energy-momentum tensor.
In this paper we will study the massless, minimally coupled scalar field, 
for which the energy-momentum tensor is
\begin{equation}
T_{\mu \nu }=\phi _{(,\mu }\phi _{,\nu) }-\frac{1}{2}g_{\mu \nu }\phi 
_{,\rho }\phi ^{,\rho } = \frac{1}{2}(\phi _{,\mu }\phi _{,\nu} +
\phi _{,\nu }\phi_{,\mu})
-\frac{1}{2}g_{\mu \nu }\phi_{,\rho }\phi ^{,\rho }. \label{eq:tmunu}
\end{equation}
The flux operator $T^{rt}= -T_{rt}$ has finite expectation values. The
product of a pair of normal ordered quadratic operators,
$\langle :\phi_{1}\phi_{2}::\phi_{3}\phi_{4}:\rangle$ can be expressed using 
Wick's theorem as
\begin{eqnarray} 
    :\phi_{1}\phi_{2}::\phi_{3}\phi_{4}: & = & 
:\phi_{1}\phi_{2}\phi_{3}\phi_{4}: \nonumber \\
& & +:\phi_{1}\phi_{3}:\langle \phi_{2}\phi_{4}\rangle_{M}+
:\phi_{1}\phi_{4}:\langle 
\phi_{2}\phi_{3}\rangle_{M} \nonumber \\ & & +:\phi_{2}\phi_{3}:\langle 
\phi_{1}\phi_{4}\rangle_{M}+
:\phi_{2}\phi_{4}:\langle \phi_{1}\phi_{3}\rangle_{M} \nonumber \\
& & +\langle 
\phi_{1}\phi_{3}\rangle_{M} \langle \phi_{2}\phi_{4}\rangle_{M}+\langle 
\phi_{1}\phi_{4}\rangle_{M} \langle \phi_{2}\phi_{3}\rangle_{M} \, .
\label{eq:} \end{eqnarray}
The first term is fully normal-ordered, the next four are 
cross terms and the final two are pure vacuum terms. The flux two-point
function for an arbitrary state can written as
\begin{eqnarray}
 &  & \langle T_{rt}(x)T_{rt}(x')\rangle \nonumber \\
 & = & \langle :T_{rt}(x)::T_{rt}(x'):\rangle \nonumber \\
 & = & \langle :T_{rt}(x)T_{rt}(x'):\rangle +\langle 
T_{rt}(x)T_{rt}(x')\rangle _{cross}+\langle 
T_{rt}(x)T_{rt}(x')\rangle _{M},\label{eq:flux2pt} 
\end{eqnarray}
where the vacuum term
\begin{equation}
\langle T_{rt}(x)T_{rt}(x')\rangle _{M}=\langle  \phi 
(x)_{,r}\phi (x')_{,r'}\rangle _{M}
\langle \phi (x)_{,t} \phi (x')_{,t'}\rangle _{M}
\end{equation}
is the expectation value in the Minkowski vacuum state and
\begin{equation}
\langle T_{rt}(x)T_{rt}(x')\rangle _{cross}=\langle : 
\phi (x)_{,r} \phi (x')_{,r'}:\rangle 
\langle \phi (x)_{,t} \phi(x')_{,t'}\rangle _{M}+\langle : \phi 
(x)_{,t} \phi (x')_{,t'}:\rangle 
\langle  \phi (x)_{,r} \phi (x')_{,r'}\rangle _{M}
\end{equation}
is a state-dependent cross term.
In the coincidence limit, $x' \rightarrow x$, 
$\langle T_{rt}(x)T_{rt}(x')\rangle _{M}$ and 
$\langle T_{rt}(x)T_{rt}(x')\rangle _{cross}$ are divergent. 
 We can see that although the off-diagonal 
components
of energy-momentum tensor are finite, their square is divergent
 and remains so even if \( \langle T_{rt}^{2}\rangle_{M}\) is subtracted.
We have cross terms  which contain state-dependent divergences. If we wish 
to give meaning 
to the notion of the squared flux as a local quantity, it is necessary to 
remove this divergence. One possibility is simply to remove both the vacuum and 
cross term and 
define \(\langle T_{rt}^{2}\rangle =\langle :T_{rt}^{2}:\rangle \).
In the following section, we will consider the normal ordered term alone,
and in Sect.~\ref{sec:cross}, we will return to the issue of whether there 
is a nontrivial physical content to the cross term.

Note that the quantities we will investigate, 
$\langle :T_{rt}(x)T_{rt}(x'):\rangle $ 
and \mbox{$\langle :T_{rt}(x)T_{rt}(x'):\rangle$}  $+\langle 
T_{rt}(x)T_{rt}(x')\rangle _{cross}$,   are distinct from the
stress tensor correlation function given by 
\begin{equation}
C_{\mu\nu\rho\sigma} = \langle T_{\mu\nu}(x) T_{\rho\sigma}(x')\rangle
- \langle T_{\mu\nu}(x)\rangle \langle T_{\rho\sigma}(x')\rangle \,. 
                                \label{eq:corr_fnt}
\end{equation}
This latter quantity is independent of the choice of renormalization
in the sense that it is unchanged by subtracting a c-number from $T_{\mu\nu}$,
but it is singular in the coincidence limit. This correlation function was
used by Muller and Schmid \cite{MS} to discuss cosmological perturbations
due to quantum fields and by Carlitz and Willey \cite{CW} in the context of
black hole evaporation. However, for the questions which we wish to pose, it 
seems more natural to examine the various terms in Eq.~(\ref{eq:flux2pt}).

\section{Flux fluctuations using the  normal-ordering scheme }
\label{sec:FFno}

\subsection{Formula for \protect\( \langle :T^{2}_{rt}:\rangle 
\protect \)}

Consider the massless scalar field, whose stress tensor is given by 
Eq.~(\ref{eq:tmunu}).
The normal-ordered expectation value of a product of fluxes  is shown 
in Appendix A to be 

\begin{eqnarray}
\langle :T_{rt}(x)T_{rt}(x'):\rangle  & = & \langle 
:T_{rt}(x):\rangle \langle :T_{rt}(x'):\rangle +\langle :\phi 
_{,r}(x)\phi _{,r}(x'):\rangle \langle :\phi _{,t}(x)\phi 
_{,t}(x'):\rangle \nonumber \label{math-nonumber} \\
 &  & +\langle :\phi _{,r}(x)\phi _{,t}(x'):\rangle \langle :\phi 
_{,t}(x)\phi _{,r}(x'):\rangle \label{eq:fluxcorr} 
\end{eqnarray}
and the squared flux is 

\begin{equation}
\langle :T_{rt}^{2}(x):\rangle =lim_{x'\rightarrow x}\langle 
:T_{rt}(x)T_{rt}(x'):\rangle 
\end{equation}

\subsection{Two dimensional moving mirror}
\label{sec:2dmirror}

In flat space-time, boundaries induce vacuum energy and stress. If 
the boundaries
move, then particles can be created. A single reflecting boundary 
(mirror) can
create particles if it undergoes non-uniform acceleration.

We follow the treatment of Fulling and Davies \cite{FD76,DF77}, and consider a 
massless
scalar field in two dimensional flat spacetime with an arbitrary mirror 
trajectory,
as illustrated in Fig. 1.

\begin{figure}
\begin{center}
\leavevmode\epsfysize=6.5cm\epsffile{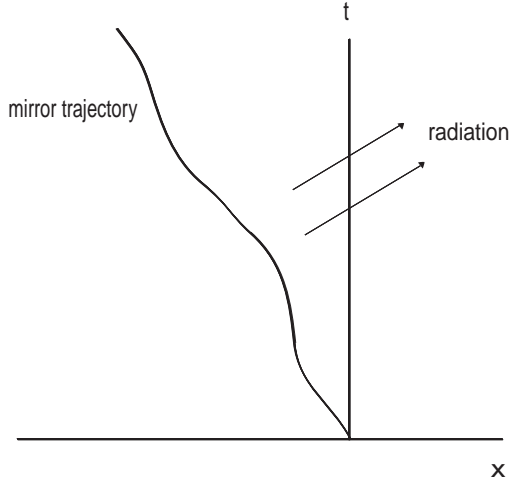}
\end{center}
\caption{
The worldline of a mirror moving to the left. The radiation 
emitted to the right is illustrated.}
\label{fig:mirror}
\end{figure}

The trajectory for the mirror is

\begin{equation}
x=z(t),
\end{equation}
where \( |z'(t)|<1 \), and \( z=0 \) for \( t\leq 0 \).
The boundary condition for the scalar field \( \phi  \) is

\begin{equation}
\phi (t,z(t))=0,
\end{equation}
and the positive frequency mode function for \( t>0 \) is given by

\begin{equation}
\phi _{\omega }(x)=i(4\pi \omega )^{-1/2}[e^{-i\omega v}-e^{-i\omega 
(2\tau _{u}-u)}].
\end{equation}
Here \( u\equiv t-x \) and \( v\equiv t+x \) are null coordinates, 
and the
parameter \( \tau _{u} \) is defined by

\begin{equation}
\tau _{u}-z(\tau _{u})=u.
\end{equation}
The phase of the reflected wave is a function of \( u \) only and is 
defined
by

\begin{equation}
p(u)=2\tau _{u}-u.
\end{equation}
The phase change of the out-going mode is due to the Doppler shift at 
the moving
mirror. It is not surprising to see that the moving mirror can create 
particles,
since the mirror boundary condition changes a positive frequency mode 
into a
mixture of positive and negative frequencies.

The quantum field operator can be written as

\begin{equation}
\phi (x,t)=\int _{0}^{\infty }d\omega [a_{\omega }\phi _{\omega 
}+a_{\omega }^{\dagger }\phi _{\omega }^{*}],
\end{equation}
where \( a_{\omega } \) and \( a_{\omega }^{\dagger } \) are 
annihilation
and creation operators, respectively. The in-vacuum state \( 
|0\rangle  \) is
defined by

\begin{equation}
\begin{array}{cc}
a_{\omega }|0\rangle =0, & \forall 
\end{array}\omega .
\end{equation}
We use the point-splitting method to extract out the Minkowski vacuum 
divergence.
Let \( \phi ^{*}_{\omega} \) be replaced by \( \phi ^{*}_{\omega}(t+\epsilon 
,x) \),  
and 
\begin{equation}
\langle \phi _{,\mu }\phi _{,\nu }\rangle =\lim _{\epsilon 
\rightarrow 0}\int _{0}^{\infty }\phi _{\omega ,\mu }(r,t)\phi 
^{*}_{\omega ,\nu }(r,t+\epsilon )d\omega . \label{eq:2pt}
\end{equation}
The derivatives of the mode functions become

\begin{eqnarray}
\phi _{\omega ,t} & = & \left(\frac{\omega }{4\pi }\right)^{1/2}[e^{-i\omega 
v}-p'(u)e^{-i\omega p(u)}],\\
\phi _{\omega ,r} & = & \left(\frac{\omega }{4\pi }\right)^{1/2}[e^{-i\omega 
v}+p'(u)e^{-i\omega p(u)}],\\
\phi _{\omega ,t}^{*} & = & \left(\frac{\omega }{4\pi }\right)^{1/2}[e^{i\omega 
(v+\epsilon )}-p'(u+\epsilon )e^{i\omega p(u+\epsilon )}],
\end{eqnarray}
and

\begin{equation}
\phi _{\omega ,r}^{*}=\left(\frac{\omega }{4\pi }\right)^{1/2}[e^{i\omega 
(v+\epsilon )}+p'(u+\epsilon )e^{i\omega p(u+\epsilon )}].
\end{equation}
We insert these expressions into Eq.~(\ref{eq:2pt}), and evaluate the \( 
\omega  \)-integration
using

\begin{equation}
\int _{0}^{\infty }e^{ibx}dx=\frac{i}{b},
\end{equation}
and

\begin{equation}
\int _{0}^{\infty }\omega e^{ia\omega }d\omega =\frac{\partial 
}{i\partial a}\int _{0}^{\infty }e^{ia\omega }d\omega 
=-\frac{1}{a^{2}}.
\end{equation}
Expanding the results in a power series in \( \epsilon  \) yields

\begin{eqnarray}
\langle \phi _{,r}\phi _{,t}\rangle  & = & -\frac{1}{4\pi }\left[ 
\frac{1}{4}\left(\frac{p''}{p'}\right)^{2}-\frac{1}{6}\frac{p'''}{p'}\right] 
+O(\epsilon ),\\
\langle \phi _{,t}\phi _{,r}\rangle  & = & -\frac{1}{4\pi }\left[ 
\frac{1}{4}\left(\frac{p''}{p'}\right)^{2}-\frac{1}{6}\frac{p'''}{p'}\right] 
+O(\epsilon ),\\
\langle \phi _{,r}\phi _{,r}\rangle  & = & -\frac{1}{4\pi }\left[ 
\frac{2}{\epsilon 
^{2}}+\frac{2p'}{(v-p(u))^{2}}-\frac{1}{4}\left(\frac{p''}{p'}\right)^{2}+
\frac{1}{6}\frac{p'''}{p'}\right] ,
\end{eqnarray}
and

\begin{equation}
    \langle \phi_{,t}\phi_{t}\rangle =-\frac{1}{4\pi }\left[ 
    \frac{2}{\epsilon ^{2}}-\frac{2p'}{(v-p(u))^{2}}
    -\frac{1}{4}\left(\frac{p''}{p'}\right)^{2}+
\frac{1}{6}\frac{p'''}{p'}\right].    
\end{equation}

Here \( \langle \phi _{,r}\phi _{,t}\rangle  \) and \( \langle \phi 
_{,t}\phi _{,r}\rangle  \)
are finite, and \( \langle \phi _{,r}\phi _{,r}\rangle  \)and \( 
\langle \phi _{,t}\phi _{,t}\rangle  \)
can be renormalized by discarding the \( \epsilon ^{-2} \) term. This 
term
is independent of \( p(u) \) (i.e. independent of trajectory) and can 
be recognized
as the Minkowski vacuum divergence. The normal-ordered operator 
products are

\begin{eqnarray}
\langle :\phi _{,r}\phi _{,t}:\rangle  & = & \langle \phi _{,r}\phi 
_{,t}\rangle , \\
\langle :\phi _{,t}\phi _{,r}:\rangle  & = & \langle \phi _{,t}\phi 
_{,r}\rangle ,\\
\langle :\phi _{,r}\phi _{,r}:\rangle  & = & \langle \phi _{,r}\phi 
_{,r}\rangle +\frac{1}{2\pi \epsilon ^{2}},
\end{eqnarray}
and

\begin{equation}
\langle :\phi _{,t}\phi _{,t}:\rangle =\langle \phi _{,t}\phi 
_{,t}\rangle +\frac{1}{2\pi \epsilon ^{2}}.
\end{equation}
We next substitute the above relations into Eq.~(\ref{eq:fluxcorr}). 
The squared
flux for an arbitrary trajectory becomes
\begin{equation}
\langle :T_{rt}^{2}:\rangle =\left(\frac{1}{4\pi }\right)^{2}\left[ 
\frac{-4(p')^{2}}{(v-p(u))^{4}}+\frac{3}{16}\left(\frac{p''}{p'}\right)^{4}-
\frac{1}{4}
\frac{(p'')^{2}p'''}{(p')^{3}}+\frac{(p''')^{2}}{12(p')^{2}}\right] ,
   \label{eq:sqflux}
\end{equation}
and the flux is 

\begin{equation}
    \langle T^{rt}\rangle =\frac{1}{4\pi }\left[ 
\frac{1}{4}\left(\frac{p''}{p'}\right)^{2}-\frac{1}{6}\frac{p'''}{p'}\right] .   
\end{equation}

\subsubsection{A trajectory which produce a thermal spectrum of 
particles}

Fulling and Davies \cite{FD76,DF77} have discussed a particular
 mirror trajectory which produces a steady thermal flux of particles at late 
times, and which models a two-dimensional evaporating black hole.
Carlitz and Willey \cite{CW} have shown that the correlation function $C$
defined in Eq.~(\ref{eq:corr_fnt}) in this case is just that for a thermal
state. The trajectory has the asymptotic form

\begin{equation}
\begin{array}{cc}
z(t)\sim -t-Ae^{-2\kappa t+B}, & t\rightarrow \infty , \label{eq:thermal_traj}
\end{array}
\end{equation}
where \( A,B, \) and \( \kappa  \) are positive constants.  Here

\begin{equation}
p(u)=2\tau _{u}-u\sim B-A^{-\kappa (u+B)}, \quad u\rightarrow \infty .
\end{equation}
Substituting this form Eq.~(\ref{eq:sqflux}) gives the normal-ordered 
squared
flux

\begin{equation}
\langle :T_{rt}^{2}(x):\rangle =
\left(\frac{1}{4\pi }\right)^{2}\left[\frac{\kappa 
^{4}}{48}-\frac{16A^{2}\kappa ^{2}e^{- 2\kappa 
(u+B)}}{(v-B+Ae^{-\kappa (u+B)})^{4}}\right]\to \left(\frac{1}{4\pi 
}\right)^{2}\frac{\kappa ^{4}}{48}, \quad u\to \infty .
\end{equation}
Similarly, the expectation value of the flux is

\begin{equation}
\langle :T^{rt}(x):\rangle =-\langle :T_{rt}(x):\rangle =
\frac{\kappa^2}{48\pi} .
\end{equation}
The squared flux is related to the mean flux by

\begin{equation}
\langle :T_{rt}^{2}(x):\rangle =3\langle :T_{rt}:\rangle ^{2},
\end{equation}
and the relative deviation is

\begin{equation}
\frac{\langle :\triangle T_{rt}:\rangle }{\langle :T_{rt}:\rangle 
}=\frac{\sqrt{\langle :T_{rt}^{2}(x):\rangle -\langle :T_{rt}:\rangle 
^{2}}}{\langle :T_{rt}:\rangle }=\sqrt{2}.
\end{equation}
The fractional flux fluctuations are thus of order unity.

\subsubsection*{Correlation function }

The function \( \langle T_{rt}(x)T_{rt}(x')\rangle  \) is finite 
except in
the short distance limit \( x'-x\rightarrow 0 \). On the other hand, 
the normal-ordered
function \( \langle :T_{rt}(x)T_{rt}(x'):\rangle  \) has a finite value 
in this
limit. Here we restrict our discussion to the latter function.
We define a normalized correlation function as (note that this is
distinct from the function $C$ defined in Eq.~(\ref{eq:corr_fnt}) )

\begin{equation}
\xi (\triangle t)=\frac{\langle :T_{rt}(t)T_{rt}(t'):\rangle -\langle 
:T_{rt}(t):\rangle \langle :T_{rt}(t'):\rangle }{\langle 
:T_{rt}(t):\rangle \langle :T_{rt}(t'):\rangle }, \label{eq:norm_corr}
\end{equation}
where \(\triangle t=|t-t'|\) and the spatial points are coincident.
For the specific trajectory of Eq.~(\ref{eq:thermal_traj}), 
\( \xi (\triangle t) \) is

\begin{equation}
\xi (\triangle t)=288\left[\frac{e^{-\kappa \triangle t}}{(1-e^{-\kappa 
\triangle t})^{2}}-\frac{1}{(\kappa \triangle t)^{2}}\right]^{2},
\end{equation}
which is plotted in Fig. \( 2 \). Note that \(\xi (\triangle t)\) is finite 
for all \(\triangle t \ge 0\).

\begin{figure}
\begin{center}
\leavevmode\epsfysize=6.5cm\epsffile{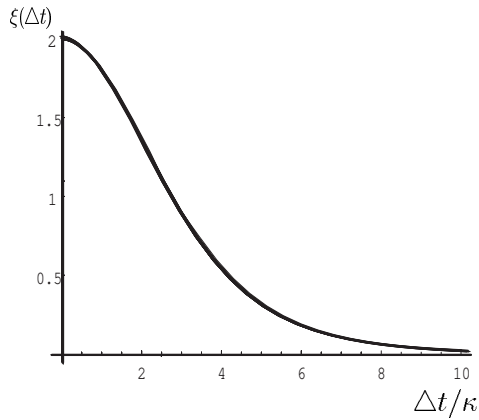}
\end{center}
\caption{
The correlation function for thermal radiation in two dimensional spacetime.}
\label{fig:cofun2d}
\end{figure}

\begin{equation}
\xi (\tau _{c})=\frac{1}{2}\xi (0),
\end{equation}
and is approximately  
\begin{equation}
\tau _{c}\approx \frac{3}{\kappa }.
\end{equation}

\subsection{Two dimensional black hole}

Fulling and Davies have shown that the mirror trajectory of 
Eq.~(\ref{eq:thermal_traj})
produces the same quantum state in the asymptotic region as does a two 
dimensional
evaporating black hole of mass \( M \) if \( \kappa =1/4M \). Thus 
the flux
and squared flux for the 2-D black hole are

\begin{equation}
\langle :T^{rt}(x):\rangle =\frac{1}{768\pi M^{2}},
\end{equation}
and

\begin{equation}
\langle :T_{rt}^{2}(x):\rangle =3\langle :T_{rt}(x):\rangle ^{2}.
\end{equation}
The correlation time \( \tau _{c} \) becomes

\begin{equation}
\tau _{c}=11M.
\end{equation}
Thus the Hawking flux undergoes large fluctuations, varying by a factor 
of order
unity on a time scale of order \( 11M \).\\

\subsection{Four dimensional black hole}
\label{sec:4dBH}

In four dimensions, the treatment of black hole evaporation becomes 
more complicated
than in two dimensions. This is due to the angular degrees of freedom 
and the
resultant potential barrier around the black hole. Ingoing and 
outgoing waves
experience scattering off of this potential barrier. We will consider 
the case
of a nonrotating, uncharged (Schwarzschild) black hole, and will 
follow the
treatment of DeWitt \cite{DeWitt}. The mode functions are of the form

\begin{equation}
u=\frac{1}{2\pi r\sqrt{2\omega }}R_{l}(\omega |r)Y_{lm}(\theta, \phi) 
\, e^{-i\omega t},
\end{equation}
where the $Y_{lm}(\theta, \phi)$ are the usual spherical harmonics.
Vector signs will be used to indicate the two independent modes which 
have the asymptotic
forms

\begin{equation}
\overrightarrow{R}_{l}(\omega |r)\to \left\{ \begin{array}{cc}
e^{i\omega r^{*}}+\overrightarrow{A}_{l}(\omega )e^{i\omega r*} & 
,r^{*}\to -\infty \\
B_{l}(\omega )e^{i\omega r^{*}} & ,r^{*}\to \infty 
\end{array}\right. 
\end{equation}
and

\begin{equation}
\overleftarrow{R}_{l}(\omega |r)\to \left\{ \begin{array}{cc}
B_{l}(\omega )e^{-i\omega r^{*}} & ,r^{*}\to -\infty \\
e^{-i\omega r^{*}}+\overleftarrow{A}_{l}(\omega )e^{i\omega r^{*}} & 
r^{*}\to \infty 
\end{array}\right. ,
\end{equation}
where \( r^{*} \) is the usual tortoise coordinate
\begin{equation}
r^{*}=r+2M\ln \left( \frac{r}{2M}-1\right) .
\end{equation}
The transmission coefficients \( B_{l} \) and reflection coefficients 
\( A_{l} \)
satisfy the relations

\begin{equation}
|\overleftarrow{A_{l}}(\omega )|=|\overrightarrow{A_{l}}(\omega )|,
\end{equation}
 
\begin{equation}
1-|\overrightarrow{A_{l}}(\omega 
)|^{2}=1-|\overleftarrow{A_{l}}(\omega )|^{2}=|B_{l}(\omega )|^{2},
\end{equation}
and 
\begin{equation}
\overrightarrow{A_{l}}^{*}(\omega )B_{l}(\omega 
)=-B^{*}_{l}\overleftarrow{A_{l}}(\omega ).
\end{equation}
The components of the energy-momentum tensor in the Unruh vacuum state near
future null infinity are of the form (See Ref. 12 for details.)
\begin{equation}
\langle \phi _{,\mu }\phi _{,\nu }\rangle \sim \sum _{l,m}\int 
_{0}^{\infty }[\overleftarrow{u}_{,\mu }\overleftarrow{u}_{,\nu 
}^{*}+ \coth(4\pi M\omega )\overrightarrow{u}_{,\mu 
}\overrightarrow{u}_{,\nu }^{*}]d\omega .  \label{eq:2pt_4d}
\end{equation}
The asymptotic form of the mode functions are 
\begin{equation}
\overrightarrow{u}=\frac{1}{2\pi r\sqrt{2\omega }}B_{l}(\omega 
)e^{i\omega r^{*}}Y_{lm}\, e^{-i\omega t},
\end{equation}
and 
\begin{equation}
\overleftarrow{u}=\frac{1}{2\pi r\sqrt{2\omega }}Y_{lm} 
e^{-i\omega t}(e^{-i\omega r^{*}}+\overleftarrow{A}_{l}(\omega 
)e^{i\omega r^{*}}).
\end{equation}
The derivatives of these mode functions become
\begin{eqnarray}
 \overleftarrow{u}_{,r} \overleftarrow{u}^{*}_{,t} & = & \frac{\omega 
|Y_{lm}|^{2}}{8\pi 
^{2}r^{2}}(1-|\overleftarrow{A_{l}}|^{2}+\overleftarrow{A_{l}}^{*}e^{-2i\omega 
r}-\overleftarrow{A_{l}}e^{2i\omega r}), \\
 \overleftarrow{u}_{,t} \overleftarrow{u}^{*}_{,r} & = & \frac{\omega 
|Y_{lm}|^{2}}{8\pi 
^{2}r^{2}}(1-|\overleftarrow{A_{l}}|^{2}-\overleftarrow{A_{l}}^{*}e^{-2i\omega 
r}+\overleftarrow{A_{l}}e^{2i\omega r}),\\
 \overleftarrow{u}_{,t} \overleftarrow{u}^{*}_{,t} & = & \frac{\omega 
|Y_{lm}|^{2}}{8\pi 
^{2}r^{2}}(1+|\overleftarrow{A_{l}}|^{2}+\overleftarrow{A_{l}}^{*}e^{-2i\omega 
r}+\overleftarrow{A_{l}}e^{2i\omega r}),\\
 \overleftarrow{u}_{,r} 
\overleftarrow{u}^{*}_{,r} & = & \frac{\omega 
|Y_{lm}|^{2}}{8\pi 
^{2}r^{2}}(1+|\overleftarrow{A_{l}}|^{2}+\overleftarrow{A_{l}}^{*}e^{-2i\omega 
r}+\overleftarrow{A_{l}}e^{2i\omega r}),\\
\overrightarrow{u}_{,r} 
\overrightarrow{u}^{*}_{,t} & = & \frac{-\omega 
|B_{l}|^{2}|Y_{lm}|^{2}}{8\pi ^{2}r^{2}},\\
\overrightarrow{u}_{,t} 
\overrightarrow{u}^{*}_{,r}& = & \frac{-\omega 
|B_{l}|^{2}|Y_{lm}|^{2}}{8\pi ^{2}r^{2}},\\
 \overrightarrow{u}_{,r} 
\overrightarrow{u}^{*}_{,r} & = & \frac{\omega 
|B_{l}|^{2}|Y_{lm}|^{2}}{8\pi ^{2}r^{2}},
\end{eqnarray}
and 
\begin{equation}
\overrightarrow{u}_{,t}
\overrightarrow{u}^{*}_{,t}=\frac{\omega 
|B_{l}|^{2}|Y_{lm}|^{2}}{8\pi ^{2}r^{2}}.
\end{equation}
Substitution of these relation into Eq.~(\ref{eq:2pt_4d}) and use of the 
summation formula
\begin{equation}
\sum _{m}|Y_{lm}|^{2}=\frac{2l+1}{4\pi },
\end{equation}
yields 
\begin{eqnarray}
\langle \phi _{,r}\phi _{,t}\rangle  & = & \sum _{l,m}\int 
_{0}^{\infty 
}[\overleftarrow{u}_{,r}\overleftarrow{u}_{,t}^{*}+ \coth(4\pi M\omega 
)\overrightarrow{u}_{,r}\overrightarrow{u}_{,t}^{*}]d\omega \nonumber 
\\
 & = & \frac{-\sin \theta }{4\pi ^{2}}\sum _{l,m}|Y_{lm}|^{2}\int 
_{0}^{\infty }\frac{|B_{l}|^{2}\omega }{e^{8\pi M\omega }-1}d\omega 
\nonumber \\
 &  & +\frac{\sin \theta }{32\pi ^{3}}\sum _{l}(2l+1)\int 
_{0}^{\infty }\omega (\overleftarrow{A_{l}}^{*}e^{-2i\omega 
r}-\overleftarrow{A_{l}}e^{2i\omega r})d\omega . 
\end{eqnarray}
The first term on the right hand side is \( \langle T_{rt}\rangle  \). Similar 
calculations give
us these derivatives in terms of the mean flux \( \langle 
T^{rt}\rangle  \)
as 
\begin{eqnarray}
\langle \phi _{,r}\phi _{,t}\rangle  & = & -\langle T^{rt}\rangle 
+I_{2,}\\
\langle \phi _{,t}\phi _{,r}\rangle  & = & -\langle T^{tr}\rangle 
-I_{2,} \\
\langle \phi _{,r}\phi _{,r}\rangle  & = & \langle T^{rt}\rangle 
+I_{0}-I_{1,}
\end{eqnarray}
and 
\begin{equation}
\langle \phi _{,t}\phi _{,t}\rangle = \langle T^{rt}\rangle 
+I_{0}+I_{1}.
\end{equation}
where the integrals \( I_{0}, \) \( I_{1} \) and \( I_{2} \) are 
\begin{equation}
I_{0}=\frac{\sin \theta }{16\pi ^{3}}\sum _{l}(2l+1)\int ^{\infty 
}_{0}\omega d\omega ,
\end{equation}
 
\begin{equation}
I_{1}=\frac{\sin \theta }{16\pi ^{3}}\sum _{l}(2l+1)\int ^{\infty 
}_{0}\omega A(\omega )\cos (\delta (\omega )+2r\omega )d\omega ,
\end{equation}
and 
\begin{equation}
I_{2}=-i\frac{\sin \theta }{16\pi ^{3}}\sum _{l}(2l+1)\int ^{\infty 
}_{0}\omega A(\omega )\sin (\delta (\omega )+2r\omega )d\omega .
\end{equation}
 We can identity \( I_{0} \) as the Minkowski divergence, and 
symmetrization
removes the pure imaginary term \( I_{2} \). Discarding these two 
terms yields
\begin{eqnarray}
\frac{1}{2} (\langle :\phi _{,r}\phi _{,t}:\rangle 
+ \langle :\phi _{,t}\phi _{,r}:\rangle) & = & \langle :T_{rt}:\rangle 
= \langle :T_{tr}:\rangle ,\\
\langle :\phi _{,r}\phi _{,r}:\rangle  & = & -\langle :T_{rt}:\rangle 
-I_{1,}
\end{eqnarray}
 and 
\begin{equation}
\langle :\phi _{,t}\phi _{,t}:\rangle =-\langle :T_{rt}:\rangle 
+I_{1.}
\end{equation}
Using the relations
\[
\begin{array}{cc}
A(\omega )\to A_{0,} & \omega \to 0,\\
A(\omega )\to 0, & \omega \to \infty ,
\end{array}\]
and 
\[
\begin{array}{cc}
\delta (\omega )\to \delta _{0}, & \omega \to \infty ,
\end{array}\]
the integral \( I_{1} \) at large distance, \( r\to \infty  \), 
becomes

\begin{eqnarray}
\int ^{\infty }_{0}\omega A(\omega )\cos (\delta (\omega )+2r\omega 
)d\omega  & = & \frac{1}{r^{2}}
\int^{\infty }_{0}xA\left(\frac{x}{r}\right)\cos 
\left[\delta \left(\frac{x}{r}\right)+2x\right]d\omega \nonumber  \\
 & \sim  & \frac{A_{0}}{r^{2}}\int ^{\infty }_{0}\cos (2x+\delta 
_{0})xdx\propto \frac{1}{r^{2}}.\nonumber \\
 & 
\end{eqnarray}

Thus the term \( I_{1} \) is much smaller than the mean flux 
\begin{equation}
\langle T^{rt}\rangle = \frac{\sin \theta }{16\pi ^{3}}\sum 
_{l}(2l+1)\int ^{\infty }_{0}\frac{|B_{l}|^{2}}{e^{8\pi M\omega 
-1}}d\omega 
\end{equation}
which is a nonzero constant at large distance, and hence \( I_{1} \) can be
ignored. The squared flux becomes 
\begin{eqnarray}
\langle :T_{rt}^{2}:\rangle  & = & \langle :T_{rt}:\rangle 
^{2}+\langle :\phi _{,r}\phi _{,t}:\rangle \langle :\phi _{,t}\phi 
_{,r}:\rangle +\langle :\phi _{,r}\phi _{,r}:\rangle \langle :\phi 
_{,t}\phi _{,t}:\rangle \nonumber \\
 & = & 2\langle :T_{rt}:\rangle ^{2}+(\langle :T_{rt}:\rangle 
-I_{1})(\langle :T_{rt}:\rangle +I_{1})\nonumber \\
 & \approx  & 3\langle T_{rt}\rangle ^{2}. 
\end{eqnarray}
We thus get the same relation between the squared flux and the mean 
flux as
in the case of a two dimensional evaporating black hole. 

Next we will discuss the 
normal-ordered
correlation function and ignore the contribution from \( I_{1} \). From 
Eq.~(\ref{eq:2pt_4d}), we have
\begin{eqnarray}
\langle :\phi _{,r}(x)\phi _{,r}(x'):\rangle  & = & Re\left[ 
\frac{1}{16\pi ^{2}r^{2}}\sum _{l}(2l+1)\int _{0}^{\infty 
}\frac{|B_{l}|^{2}\omega e^{i\omega \triangle t}}{e^{8\pi M\omega 
}-1}d\omega \right] ,\\
\langle :\phi _{,t}(x)\phi _{,t}(x'):\rangle  & = & Re\left[ 
\frac{1}{16\pi ^{2}r^{2}}\sum _{l}(2l+1)\int _{0}^{\infty 
}\frac{|B_{l}|^{2}\omega e^{i\omega \triangle t}}{e^{8\pi M\omega 
}-1}d\omega \right] ,\\
\langle :\phi _{,r}(x)\phi _{,t}(x'):\rangle  & = & Re\left[ 
-\frac{1}{16\pi ^{2}r^{2}}\sum _{l}(2l+1)\int _{0}^{\infty 
}\frac{|B_{l}|^{2}\omega e^{i\omega \triangle t}}{e^{8\pi M\omega 
}-1}d\omega \right] ,
\end{eqnarray}
and 
\begin{equation}
\langle :\phi _{,t}(x)\phi _{,r}(x'):\rangle =Re\left[ 
-\frac{1}{16\pi ^{2}r^{2}}\sum _{l}(2l+1)\int _{0}^{\infty 
}\frac{|B_{l}|^{2}\omega e^{i\omega \triangle t}}{e^{8\pi M\omega 
}-1}d\omega \right] ,
\end{equation}
where \( \triangle t=t'-t \). The correlation function defined in 
Eq.~(\ref{eq:norm_corr})   now becomes
\begin{equation}
\xi (\triangle t)=\frac{2\left( Re\left[ \sum _{l}(2l+1)\int 
_{0}^{\infty }\frac{|B_{l}|^{2}\omega e^{i\omega \triangle 
t}}{e^{8\pi M\omega }-1}d\omega \right] \right) ^{2}}{\left( Re\left[ 
\sum _{l}(2l+1)\int _{0}^{\infty }\frac{|B_{l}|^{2}\omega }{e^{8\pi 
M\omega }-1}d\omega \right] \right) ^{2}}. \label{eq:4dcorr}
\end{equation}
For our purposes, the transmission coefficient may be approximated by 
a step function
\begin{equation}
B_{l}(p)\approx \Theta (\sqrt{27}M\omega -l).
\end{equation}
In this approximation, modes with energies below the peak of the angular
momentum barrier are assumed to be perfectly reflected, and those with
energies above the peak are completely transmitted. This is a reasonably
good approximation, as may be seen by examining Figure 1 in Ref.~\cite{JMO},
where numerical results for the transmission and reflection coefficients 
are given.  The summation on \( l \) yields 
\begin{equation}
\sum _{l}(2l+1)|B_{l}|^{2}=\sum _{l}(2l+1)\Theta 
(\sqrt{27}M\omega -l)=27M^{2}\omega^{2}+6\sqrt{3}M\omega +1 \, ,
\end{equation}
and the numerator of Eq.~(\ref{eq:4dcorr}) becomes 
\begin{eqnarray}
S(\triangle t) & = & Re\left(\sum _{l}(2l+1)\int _{0}^{\infty 
}\frac{|B_{l}|^{2}\omega e^{i\omega \triangle t}}
{e^{8\pi M\omega}-1}d\omega\right)\nonumber \\
 & = & Re\left(\int _{0}^{\infty 
}\frac{(27M^{2}\omega^{2}+6\sqrt{3}M\omega +1)\omega e^{ip\triangle t}}{e^{8\pi 
M\omega}-1}d\omega\right)\nonumber \\
 & = & \frac{27}{(8\pi )^{4}M^{2}}Re\left(\int _{0}^{\infty 
}\frac{x^{3}e^{ibx}}{e^{x}-1}dx\right)+\frac{6\sqrt{3}}{(8\pi 
)^{3}M^{2}}Re\left(\int _{0}^{\infty 
}\frac{x^{2}e^{ibx}}{e^{x}-1}dx\right)\nonumber \\
 &  & +\frac{1}{(8\pi )^{2}M^{2}}Re\left(\int _{0}^{\infty 
}\frac{xe^{ibx}}{e^{x}-1}dx\right), \label{eq:S}
\end{eqnarray}
where \( x=8\pi M\omega \) and \( b=\triangle t/8\pi M \).
The correlation function is 
\begin{equation}
\xi (\triangle t)=\frac{2S(\triangle t)^{2}}{S(0)^{2}},
\end{equation}
and is plotted in Fig. \( 3 \).

\begin{figure}
\begin{center}
\leavevmode\epsfysize=6cm\epsffile{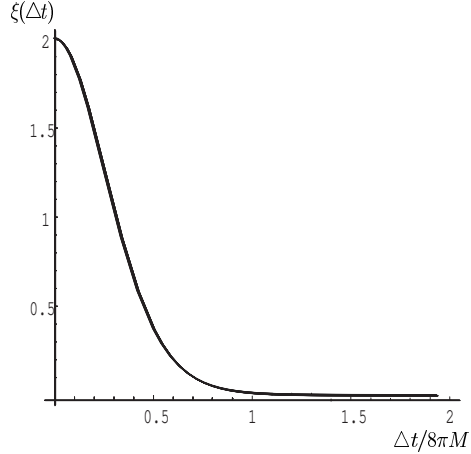}
\end{center}
\caption{
The correlation function for the radiation from a black hole in four
dimensional spacetime.}
\label{fig:cofun4d}
\end{figure}

The correlation time \( \tau _{c} \) is around \( 0.3(8\pi M)\approx 
8M \)
. As in the two dimensional case, the four dimensional Hawking 
radiation undergoes
large flux fluctuations on a time scale of about \( 8M \).

\subsection{Flux Fluctuations as Thermal Fluctuations}
\label{sec:thermalfluct}

It is reasonable to expect that the flux fluctuations computed in the previous
subsection can be interpreted as ordinary thermal fluctuations. Thermal
fluctuations of the energy in the canonical ensemble are described by the 
relation
\begin{equation}
\frac{\Delta E}{E} = \frac{T\, \sqrt{k_B C_V}}{E} \, ,
\end{equation}
where $E$ is the mean energy at temperature $T$ for a system with heat
capacity $C_V$. In the case of thermal radiation, $E \propto T^4$, so
$C_V = 4 E/T$, and hence
\begin{equation}
\frac{\Delta E}{E} = 2 \sqrt{\frac{k_B T}{E}} \,.
\end{equation}
Note that $E/(k_B T)$ is a measure of the mean number of photons in the 
thermal radiation, so the above result is the familiar $1/\sqrt{N}$
statistical fluctuation.

Let us take $E$ to be the energy emitted by a black hole
in one correlation time, $\tau_c \approx 8 M$, and the power emitted to
be that calculated by Page \cite{Page} for photon emission from a 
Schwarzscild black hole:
\begin{equation}
P = 3.4 \times 10^{-5}\, M^{-2} \,.
\end{equation}
This leads to a rather small number of photons emitted per correlation time,
\begin{equation}
\frac{E}{k_B T} \approx 8 \times 10^{-3}
\end{equation}
and rather large fractional energy fluctuations
\begin{equation}
\frac{\Delta E}{E} \approx 23 \,.
\end{equation}

This estimate is somewhat larger than the result obtained from the normal
ordered squared flux in the previous subsection. However, it is a very rough
estimate which depends upon our choice for the energy $E$ and the collecting 
time. If we had chosen to integrate the flux for a time longer than $\tau_c$, 
the statistical fluctuations would be somewhat reduced. Also, the 
spectrum of particles emitted by a black hole is not exactly  Planckian, but 
has been filtered by the angular momentum barrier.

\section{The Physics of the Cross Term}
\label{sec:cross}

We now turn to examining the cross term between the vacuum fluctuations and
the finite, state dependent parts. Recall that this contribution is singular
in the coincidence limit, and hence does not lead to a well defined local
definition of the squared flux. If it is not to be subtracted by some
renormalization method, then it can only be given physical meaning by dealing
with time or space averages. 

\subsection{Switching Functions}

One possibility is to suppose that we operationally measure the flux with 
a model detector which has a finite response time. Suppose that the response of
our detector is described by a Lorentzian function  with characteristic width
$\tau$,
\begin{equation}
f(t)=\frac{\tau}{\pi }\frac{1}{(t-t_{0})^{2}+\tau^{2}}.
\end{equation}
The averaged squared flux becomes
\begin{equation}
 \langle T_{rt}^{2}(x)\rangle _{average}=
\int _{-\infty }^{\infty }\int _{-\infty }^{\infty }f(t)f(t')\langle 
T_{rt}(x)T_{rt}(x')\rangle dtdt'.
\end{equation}
We will examine the case of the thermal flux from a mirror or black hole in
two dimensions, and assume that the sampling time is short compared to the
correlation time, $\tau \ll \tau_c$. In this case, the correlation functions
are approximately constant, and the average of the normal ordered term is 
\begin{equation}
A_2 =\int _{-\infty }^{\infty }\int _{-\infty }^{\infty }f(t)f(t')\langle 
:T_{rt}(x)T_{rt}(x'):\rangle dtdt'\approx \langle 
:T_{rt}(x)T_{rt}(x'):\rangle .
\end{equation}
The average  of the cross term can be written as
\begin{eqnarray}
 & A_1 & = \int_{-\infty}^{\infty}\int_{-\infty }^{\infty} 
f(t)f(t')\left[ \langle :\phi (x)_{,r}
\phi (x')_{,r'}:\rangle \langle 
 \phi (x)_{,t}
 \phi (x')_{,t'}\rangle_{M}+ \right. \nonumber \\
 & & \left. \langle : \phi (x)_{,t}
\phi (x')_{,t'}:\rangle 
\langle  \phi (x)_{,r}
\phi (x')_{,r'}\rangle_{M}\right] dtdt'
  \approx   2\langle :T^{rt}:\rangle \int _{-\infty}^{\infty}
f(t)f(t')\frac{1}{-2\pi (t-t')^{2}}dtdt' \nonumber \\
 & \approx  & \frac{1}{4 \pi \tau^{2} }\, \langle T^{rt}\rangle , 
\end{eqnarray}
where we have used Eq.~(\ref{eq:App_B1}) in Appendix B.

For comparison, it is of interest to give the average  of the vacuum term
\begin{eqnarray}
 &A_0 & = \int _{-\infty }^{\infty }\int _{-\infty }^{\infty 
}f(t)f(t')\langle  \phi (x)_{,r} 
\phi (x')_{,r'}\rangle _{M}\langle  \phi 
(x)_{,t}\phi (x')_{,t'}\rangle 
_{M}dtdt'\nonumber \\
 & = & \int _{-\infty }^{\infty }\int _{-\infty }^{\infty 
}f(t)f(t')\frac{1}{4\pi ^{2}(t-t')^{4}}dtdt'
  = \frac{1}{64 \pi^2 \tau^{4}} \, ,
\end{eqnarray}
using Eq.~(\ref{eq:App_B2}).

In the case of a two-dimensional black hole, where $\langle T^{rt}\rangle
=1/(768 \pi M^2)$ and $\langle :T_{rt}^2:\rangle =3\langle T_{rt}\rangle^{2}$,
we see that $A_0 \gg A_1 \gg A_2$ when $T \gg M$. If we were to let 
$T \approx M$, our assumption that the correlation functions are constant
would no longer be exact. Nonetheless, this calculation should give a 
resonable order of magnitude estimate, and predicts that all three terms
are of the same order of magnitude. 

\subsection{A Mirror as a Flux Detector}
\label{sec:mirror}

Here we will examine a model in which the flux is measured by the force
which it exerts on a reflecting or absorbing surface. All of our discussion
will be in one spatial dimension, so the  force  on a partially reflecting
surface  is 

\begin{equation}
F= r\, :T^{rt}: \,,
\end{equation}
where $0< r \leq 2$ and $\frac{1}{2} r$ is the fraction of the radiation 
which is reflected. 
Consider a mirror  with mass \( m \) which starts at rest at time $t=0$ . 
The mean velocity and mean squared velocity at $t=\tau$ are
\begin{equation}
\langle v\rangle =\frac{1}{m}\int_{0}^{\tau} \langle F\rangle dt
\end{equation}
and 
\begin{equation}
\langle v^{2}\rangle =\frac{1}{m^{2}}\int _{0}^{\tau}\int 
_{0}^{\tau} \langle F(t)F(t')\rangle dtdt'.
\end{equation}
The force two-point  function is
\begin{eqnarray}
\langle F(t)F(t')\rangle  & = & r^{2}\langle :T_{rt}(t):: T_{rt}(t'): \rangle 
                                                           \nonumber  \\
 & = &r^{2} \langle :T_{rt}(t) T_{rt}(t'): \rangle + r^{2}\langle 
T_{rt}T_{rt}\rangle _{cross}+ r^{2}\langle T_{rt} T_{rt}\rangle 
_{M}.\nonumber  \\
 &  & 
\end{eqnarray}
We are interested in the difference between the fluctuations in a given state
and those in the Minkowski vacuum, and so drop the vacuum term. 
We then define the fluctuations by 
subtracting out the square of the mean value:
\begin{equation}
\langle \triangle T^{2}\rangle =\langle T^{2}\rangle -\langle 
T\rangle ^{2}.
\end{equation}
The normal-ordered flux fluctuation is given by
\begin{eqnarray}
\langle \triangle T_{rt}^{2}\rangle _{NO} & = & \langle 
:T_{rt}(t)T_{rt}(t'):\rangle -\langle T_{rt}(t)\rangle \langle 
T_{rt}(t')\rangle \nonumber  \\
 & = & \langle :\phi _{,r}(t)\phi _{,r}(t'):\rangle \langle :\phi 
_{,t}(t)\phi _{,t}(t'):\rangle +\langle :\phi _{,r}(t)\phi 
_{,t}(t'):\rangle \langle :\phi _{,t}(t)\phi _{,r}(t'):\rangle 
,\nonumber  \\
 &  & 
\end{eqnarray}
and the cross-term contribution is
\begin{equation}
\langle \triangle T_{rt}^{2}\rangle _{cross}=\langle :\phi 
_{,r}(t)\phi _{,r}(t'):\rangle \langle \phi _{,t}(t)\phi 
_{,t}(t')\rangle _{M}+\langle :\phi _{,t}(t)\phi _{,t}(t'):\rangle 
\langle \phi _{,r}(t)\phi _{,r}(t')\rangle _{M.}
\end{equation}
The velocity fluctuation can be written as

\begin{eqnarray}
\langle \triangle v^{2}\rangle  & = & \langle v^{2}\rangle -\langle 
v\rangle ^{2}\nonumber  \\
 & = & \langle \triangle v^{2}\rangle _{NO}+\langle v^{2}\rangle 
_{cross}\nonumber  \\
 & = & \frac{r^2}{m^{2}}\int \langle \triangle T_{rt}^{2}\rangle 
_{NO}dtdt'+\frac{r^2}{m^{2}}\int \langle \triangle T_{rt}^{2}\rangle 
_{cross}dtdt'.
\end{eqnarray}

\subsubsection{Coherent state}

A coherent state describes a classical field excitation and is hence a
useful model to reveal the effects of the cross term.
Consider a single-mode coherent state \( |z\rangle  \) for a mode with
frequency $\omega_0$  
\begin{equation}
a_{\omega }|z\rangle =\delta _{\omega \omega _{0}}z|z\rangle .
\end{equation}
The free quantum field expanded in normal modes is 
\begin{equation}
\phi (x,t)=\sum _{\omega }(a_{\omega }\phi _{\omega }+a_{\omega 
}^{\dagger }\phi _{\omega }^{*}),
\end{equation}
where the mode function for a standing wave in a box of length \( L 
\) is 
\begin{equation}
\phi _{\omega }=\frac{1}{\sqrt{2\omega L}}e^{-i\omega t}\sin (\omega x).
\end{equation}
Assume that the mirror remains approximately stationary, as will be the case
when its mass is large, and set \( x=x' \). Further,  let 
\( z=Re^{i\triangle } \) and find 
\begin{equation}
\langle z|:\phi _{,r}(x)\phi _{,r'}(x'):|z\rangle =\frac{\omega 
_{0}|z|^{2}\cos ^{2}(\omega _{0}r)}{L}[\cos (\omega _{0}(t-t'))+\cos 
(2\triangle -\omega _{0}(t+t'))],
\end{equation}
and
\begin{equation}
\langle z|:\phi _{,t}(x)\phi _{,t'}(x'):|z\rangle =\frac{\omega 
_{0}|z|^{2}\sin ^{2}(\omega _{0}r)}{L}[\cos (\omega _{0}(t-t'))-\cos 
(2\triangle -2\omega _{0}t+\omega _{0}(t+t'))].
\end{equation}
We also have that
\begin{equation}
\langle  \phi (x)_{,t} \phi 
(x')_{,t'}\rangle _{M}=\langle  \phi 
(x)_{,r}\phi (x')_{,r'}\rangle 
_{M}=-\frac{1}{2\pi (t-t')^{2}}.
\end{equation}
For the coherent state, the only fluctuations come from the cross 
term because
\begin{equation}
\langle z|:T_{\mu \nu }T_{\rho \sigma }:|z\rangle =\langle z|:T_{\mu 
\nu }:|z\rangle \langle z|:T_{\rho \sigma }:|z\rangle .
\end{equation}
The velocity fluctuation is then
\begin{eqnarray}
\langle \triangle v^{2}\rangle  & = & \frac{r^{2}}{m^{2}}\int \langle 
\triangle T_{rt}^{2}\rangle _{cross}dtdt'\nonumber \\
 & = & \frac{-\omega _{0}|z|^{2}r^{2}}{\pi Lm^{2}}\int \{\cos ^{2}(\omega 
_{0}r)[\cos (\omega _{0}(t-t'))+\cos (2\triangle -\omega 
_{0}(t+t'))]\nonumber \\
 &  & +\sin ^{2}(\omega _{0}r)[\cos (\omega _{0}(t-t'))-\cos 
(2\triangle -2\omega _{0}t+\omega 
_{0}(t+t'))]\}\frac{1}{(t-t')^{2}}dtdt'.\nonumber  
\\
 &  & 
\end{eqnarray}

This integral is poorly defined due to the singularity of the integrand at
$t=t'$. A possible resolution of this difficulty is to employ a trick
which has been used by various authors under the rubrics of ``generalized
principle value'' \cite{Davies} or ``differential regularization'' \cite{FJL}.
In any case, it involves writing the singular factor as a derivative of
a less singular function, and then integrating by parts. We will also assume 
that the flux is adiabatically switched on in the past and off again in the
future, so that any surface terms vanish. Thus we may use relations such as
\begin{equation}
\int f(t,t')\frac{1}{(t-t')^{2}} dt dt'=\frac{1}{2}\int 
[\partial_t \partial_{t'} f(t,t')] \,\ln ((t-t')^{2})  dt dt' \,.
\end{equation}
Let \( r=0 \) , \( x=\omega _{0}t \), and \( x'=\omega _{0}t' \). The 
velocity fluctuation becomes
\begin{equation}
\langle \triangle v^{2}\rangle   =  -\frac{r^2 \omega 
_{0}^{3}|z|^{2}}{2\pi Lm^{2}}\int _{0}^{\tau}\int _{0}^{\tau}[\cos(\omega 
_{0}t-\omega _{0}t')+ \cos(2\triangle -\omega _{0}t-\omega _{0}t')]\ln 
[\omega _{0}(t-t')]^2 dtdt'
\end{equation}
The integral may now be written in terms of the variables $U=x-x'$ and
$V=x+x'$, using the identity
\begin{equation}
\int _{0}^{\tau}\int _{0}^{\tau}dt dt'=\frac{1}{2}\left(\int _{-\tau}^{0}dU\int 
_{-U}^{U+2\tau}dV+\int _{0}^{\tau}dU\int _{U}^{2\tau -U}dV\right), 
\label{eq:changevarr}
\end{equation}
and evaluated in terms of sine and cosine integral functions. We are primarily
interested in the asymptotic form for large $\tau$, which is
\begin{equation}
\langle \triangle v^{2}\rangle \sim 
 \frac{2 r^2 \omega _{0}^{2}|z|^{2}}{3Lm^{2}}\, \tau  \,.
\end{equation}

   This result shows that the cross term leads to a contribution to the
mean squared velocity of the mirror which grows linearly in time. This
is the characteristic time dependence of a random walk process. It is useful
to compare the fluctuations with the mean velocity
\begin{eqnarray}
\langle v\rangle  & = & \frac{r}{m}\int _{0}^{\tau} 
\langle T^{rt}\rangle dt\nonumber \\
 & = & \frac{|z|^{2}\, r}{4Lm}\sin(2\omega_{0}x)
[\cos(2\triangle)-\cos(2\triangle -2\omega_{0} \tau )]. 
\end{eqnarray}
The mean velocity happens to vanish at the special point $x=0$ at which we 
evaluated $\langle \triangle v^{2}\rangle$. However, at a more general point
it is of order
\begin{equation}
\langle :v:\rangle \approx \frac{|z|^{2}\, r}{Lm},
\end{equation}
and the fractional velocity fluctuations becomes of order 
\begin{equation}
\frac{\sqrt{\langle \triangle v^{2}\rangle }}{\langle v\rangle}\approx 
\frac{\omega _{0} \sqrt{L \tau}}{ |z|}.
\end{equation}
Although this quantity grows in time, it is also inversely proportional to
the amplitude $|z|$. Thus for a nearly classical state ($|z| \gg 1$), it can
remain small for a very long time.

\subsubsection{Thermal state created by moving mirror}

Consider the 2-D moving mirror with an arbitrary trajectory. These 
flux fluctuation, without the vacuum term, can be written as
\begin{eqnarray}
    & &\langle \triangle T_{rt}^{2}\rangle _{total} = 
\langle \triangle T_{rt}^{2}\rangle _{no} + 
\langle \triangle T_{rt}^{2}\rangle _{cross}  \nonumber \\
 & =&  2\left(\frac{1}{4\pi}\right)^{2}
\left[\frac{p'(u)p'(u')}{(p(u)-p(u'))^{2}}
-(\frac{p'(u')}{(v-p(u'))^{2}})^{2}-(\frac{p'(u)}{(v-p(u))^{2}})^{2}-
     \frac{1}{(t-t')^{4}}\right],    
\end{eqnarray}
where
\begin{eqnarray}
\langle \triangle T_{rt}^{2}\rangle _{no} & = & 2\left(\frac{1}{4\pi 
}\right)^{2}\left[-(\frac{p'(u')}{(v-p(u'))^{2}})^{2}-
(\frac{p'(u)}{(v-p(u))^{2}})^{2}+(\frac{p'(u)p'(u')}{(p(u)-p(u'))^{2}})^{2}+
\frac{1}{(t-t')^{4}} \right. \nonumber \\
 &  & \left. -\frac{2p'(u)p'(u')}{(p(u)-p(u'))^{2}(t-t')^{2}}\right],
\end{eqnarray}
and
\begin{equation}
\langle \triangle T_{rt}^{2}\rangle _{cross}=4\left(\frac{1}{4\pi 
}\right)^{2}\left[\frac{p'(u)p'(u')}{(p(u)-p(u'))^{2}}-
\frac{1}{(t-t')^{2}}\right] \frac{1}{(t-t')^{2}}.
\end{equation}
The total velocity fluctuation is 
\begin{eqnarray}
 \langle \triangle v^{2}\rangle_{total} & = & 
    \frac{r^{2}}{m^{2}}\int_{0}^{\tau} \int_{0}^{\tau}\langle 
\triangle T_{rt}^{2}\rangle_{total}dtdt'    \nonumber \\
 & = & \frac{2\, r^{2}}{(4\pi )^{2}m^{2}}\int_{0}^{\tau}\int_{0}^{\tau}
\left\{ \left[ 
\frac{p'(u)p'(u')}{(p(u)-p(u'))^{2}}\right]^{2} \right. \nonumber \\
&-& \left. \left[\frac{p'(u)}{(v'-p(u))^{2}}\right]^{2}
-\left[\frac{p'(u')}{(v-p(u'))^2}\right]^{2}-
\frac{1}{(t-t')^{4}}\right\} dtdt'.    \label{eq:velfluct}   
\end{eqnarray}
As before, we suppose that $ \langle \triangle T_{rt}^{2}\rangle _{total}$
is multiplied by a swithching function which vanishes in the past and in the
future. Then the $(t-t')^{-4}$ term gives no contribution after an integration
by parts. Note that this term is of the same form as the pure vacuum
contribution, so our results will not depend upon whether the vacuum part
was subtracted beforehand or not. 

Consider the trajectory which produces a thermal spectrum, 
\( p(u)=B-Ae^{-\kappa (u+B)} \) and again assume that the detector remains at
a fixed location, which we take to be $x=0$. Then $u=v=t$ and $u'=v'=t'$. The
integral of two middle terms in Eq.~(\ref{eq:velfluct}) may be shown to
approach a constant in the limit of  large sampling time \( \tau \): 
\begin{equation}
\int_{0}^{\tau}\int_{0}^{\tau}\left\{ 
\left[\frac{p'(u)}{(v'-p(u))^{2}}\right]^{2}+
\left[\frac{p'(u')}{(v-p(u'))^{2}}\right]^{2}\right\} 
dtdt'\to constant.
\end{equation}
As we will see, this is small compared to the leading term, which grows 
linearly in $\tau$.
The total velocity fluctuation is now
\begin{eqnarray}
\!\!\!\!\!\!\!\!\!\!\!\!\!\!\!\! \langle \triangle v^{2}\rangle _{total} 
& \sim &  
\frac{2\, r^{2}}{(4\pi )^{2}m^{2}}\int _{0}^{\tau}\int _{0}^{\tau}\left[ 
\frac{p'(u)p'(u')}{(p(u)-p(u'))^{2}}\right]^{2} dtdt'\nonumber \\
 & = & 
\frac{\kappa ^{4}\, r^{2}}{8\pi ^{2}m^{2}}\int_0^\tau 
\frac{e^{2\kappa (t-t')}}{(1-e^{\kappa (t-t')})^{4}}dt dt'
= \frac{\kappa ^{4}\, r^{2}}{4\pi ^{2}m^{2}} \int_0^\tau (\tau -U)\, 
\frac{e^{2\kappa U}}{(1-e^{\kappa U})^{4}}\, dU \,,
\end{eqnarray}
where in the last step we have used the change of variables given in 
Eq.~(\ref{eq:changevarr}). At late times, we have
\begin{equation}
\langle \triangle v^{2}\rangle _{total} \sim 
\frac{\tau \kappa ^{4}\, r^{2}}{4\pi ^{2}m^{2}} \int_0^\tau 
\frac{e^{2\kappa U}}{(1-e^{\kappa U})^{4}}\, dU \sim
\frac{\tau \kappa ^{3}\, r^{2}}{2\pi ^{2}m^{2}}
\int_{-\infty}^\infty \frac{e^{2x}}{(1-e^{x})^{4}}\, dx \,,
\end{equation}
where $x=\kappa U$. If we ignore the singularity in the integrand, this integral
may be evaluated directly:
\begin{equation}
\int_{-\infty}^\infty \frac{e^{2x}}{(1-e^{x})^{4}}\, dx =
-\left[\frac{3e^{x} -1}{6(1-e^{x})^{3}} \right]^{x=\infty}_{x=-\infty} 
=\frac{1}{6}
\end{equation}
A more rigorous approach is to use the relation 
\begin{equation}
\frac{1}{x^4} = -\frac{1}{12} \frac{d^4}{dx^4} \ln x^2 ,
\end{equation}
 integrate by parts, and then evaluate the
resulting integral numerically:
\begin{equation}
\int_{-\infty}^\infty \frac{1}{x^4}\, \frac{e^{2x}}{(1-e^{x})^{4}}\, dx =
-\frac{1}{12} \int_{-\infty}^\infty \ln x^2 \frac{d^4}{dx^4} \left[
\frac{x^4 e^{2x}}{(1-e^{x})^{4}} \right]\, dx \approx \frac{1}{6} \,.
\end{equation}
In either case, the result is
\begin{equation}
\langle \triangle v^{2}(\tau)\rangle _{total}=\frac{\kappa ^{3}\, r^{2}}{48\pi 
^{2}m^{2}}\tau \,. \label{eq:velfluct2}
\end{equation}
As in the case of the coherent state discussed in the previous subsection,
the mean squared velocity fluctuations grow linearly in time.

  We now wish to determine the relative contributions of the normal-ordered
and cross terms. A calculation analogous to that performed for 
$\langle \triangle v^{2}\rangle _{total}$ reveals that the normal-ordered
contribution is also  linearly growing in time:
\begin{equation}
\langle \triangle v^{2}\rangle _{no} \sim 
\frac{\tau \kappa^{3}\, r^{2}}{2\pi^{2}m^{2}} 
\int_{-\infty}^\infty \left[\frac{1}{x^4}
+ \frac{e^{2x}}{(1-e^{x})^{4}} -2 \frac{e^{x}}{x^2(1-e^{x})^{2}}\right] dx \,.
\end{equation}
In this case, the integrand is finite from the beginning, so no integration by 
parts is needed. The integral may be evaluated numerically to yield
\begin{equation}
\langle \triangle v^{2}\rangle _{no} \sim  
\frac{\kappa ^{3}\, r^{2} \tau }{192\pi^{2}m^{2}} (1.00) \,. 
\label{eq:velfluctno}
\end{equation}

The cross term contribution may be obtained as the difference of 
Eqs.~(\ref{eq:velfluct2}) and (\ref{eq:velfluctno}), 
but it is useful as a check to compute it
independently. If we follow the procedure used to find the asymptotic form
of $\langle \triangle v^{2}\rangle _{total}$, including an integration by
parts, the result is
\begin{equation}
\langle \triangle v^{2}\rangle _{cross} \sim 
\frac{ \kappa ^{3}\, r^{2}\tau}{2\pi ^{2}m^{2}} \int_{-\infty}^\infty \ln x^2 \,
\left[\frac{6}{x^4} - \frac{e^x(e^{2x}+4e^x+1)}{[1-e^{x}]^{4}} \right] dx \,.
\end{equation}
Again, the integral may be evaluated numerically with  the result
\begin{equation}
\langle \triangle v^{2}\rangle _{cross} \sim 
\frac{\kappa ^{3} \, r^{2}\tau }{66\pi^{2}m^{2}} (1.00)  \approx 
3 \langle \triangle T_{rt}^{2}\rangle _{no} \,.
\end{equation}
Thus to the accuracy of the numerical calculations, the three independently
computed pieces do indeed satisfy
\begin{equation}
\langle \triangle v^{2}\rangle _{total} = 
\langle \triangle v^{2}\rangle _{no} +
\langle \triangle v^{2}\rangle _{cross} \,.
\end{equation}
The normal-ordered term contributes 25\% of the total effect, as compared to
75\% from the cross term.

\subsubsection{Mass Fluctuations of Two Dimensional Black Holes}

We may use the above results to discuss the fluctuations in the mass of 
evaporating black holes in two dimensions. Define the {\it mass operator}
by
\begin{equation}
M(T) = M_0 - \int_0^\tau T^{rt}(t) dt \,,
\end{equation}
where $M_0$ is the initial mass at time $t=0$. The mean mass decreases
according to the semiclassical theory of gravity as
\begin{equation}
\langle M \rangle = M_0 - \int_0^\tau \langle T^{rt}(t) \rangle dt \,,
\end{equation}
However, the squared mass will undergo fluctuations:
\begin{equation}
 \langle M^2 \rangle -\langle M \rangle^2 =
\int_0^\tau \int_0^\tau [\langle T_{rt}(t) \, T_{rt}(t') \rangle - 
\langle T_{rt}(t) \rangle \langle T_{rt}(t') \rangle] dt dt' \,.
\end{equation}
Equations ~(\ref{eq:velfluct}) and (\ref{eq:velfluct2})  may be used to 
show that
\begin{equation}
\langle M^2 \rangle -\langle M \rangle^2 \sim \frac{\kappa^3\, \tau}{48 \pi^2}
\approx \frac{\tau }{1152 M_0^3} , \label{eq:massfluct}
\end{equation}
where in the last step we assumed that $\kappa =1/(4 M_0)$, which is a good
approximation in the early stages of evaporation. We may estimate the 
evaporation time $\tau_{evap}$ of a black hole by setting 
\begin{equation}
\tau _{evap} \approx \frac{M_0}{\langle T_{rt}(0) \rangle} = 768 \pi M_0^3 .
\end{equation}
If we set $\tau = \tau_{evap}$ in Eq.~(\ref{eq:massfluct}), the result is
\begin{equation}
\langle M^2 \rangle -\langle M \rangle^2 \approx \frac{2}{3 \pi} .
    \label{eq:massfluct2}
\end{equation}
Recall that we are working in Planck units, so the right hand side of
Eq.~(\ref{eq:massfluct2}) represents a mass fluctuation of the order of
the Planck mass. Even though this effect is quite small for macroscopic 
black holes, the cross term plays a significant role here. If one were
to normal order the product of flux operators above, then the right hand
sides of Eqs.~(\ref{eq:massfluct}) and (\ref{eq:massfluct2}) will be decreased
by factor of $1/4$. This leads to a thought experiment in which one could
use evaporating black holes to test the reality of the cross term. One 
prepares several black holes with the same initial mass, and then measures
the masses at some later time. If the mass fluctuation grows linearly in time
in accordance with Eq.~(\ref{eq:massfluct}) (or its four dimensional analog),
then one would have measued the effect of the cross term.

\section{Conclusion and discussion}
\label{sec:final}

We have seen that the flux of radiation from an evaporating black hole
undergoes large fluctuations on short times scales. One approach to the 
subject of flux flucuations involves the use of normal ordered products
of stress tensor operators. In this approach the squared flux is a 
finite, local quantity and the Hawking flux undergoes fluctuations of
order one on times scales of order $M$, the black hole's mass. These
fluctuations can be viewed as esentially statistical fluctuations due
to the small mean number of particle emitted by the black hole on this
time scale.

However, the subject of stress tensor fluctuations can be a subtle one,
and there is another approach in which one retains the state dependent
cross term in the product of stress tensor operators. This term is
divergent in the limit that both operators are evaluated at the same point.
Consequently, if it is present one cannot give a meaning to the local 
squared flux. It is, however, still possible to define time integrals of a
product of fluxes. These time integrals may be used to show that, at 
least in a two dimensional model, the mean square mass of a black hole
differs from the square of the mean mass by an amount which grows linearly
in time. Furthermore, the rate of growth is four times larger when the
cross term is retained as compared to the normal ordering approach.
Thus in principle, the two approaches have different observational 
consequences. Similarly, they give different predictions for the velocity
fluctuations of a material body, such as the mirror discussed in 
Sect.~\ref{sec:mirror}.

In either approach, we are dealing with thermal radiation (or filtered
radiation) in the asymptotic region far from the black hole. Thus the 
ambiguity in how to treat the fluctuations is not confined to the specific 
case of a black hole, but is present in a general quantum state. Because
we are working in the asymptotic region, we cannot directly address the
issue of horizon fluctuations caused by quantum stress tensor fluctuations
\cite{hor}. Horizon fluctuations must be far below the Planck scale in order
that Hawking's semiclassical derivation \cite{Hawking} of black hole 
evaporation hold. Estimates of the scale of the scale of the horizon
fluctuations due to quantization of the gravitational field (``active'' as
opposed to ``passive'' fluctuations) indicate that Hawking's derivation does 
indeed hold for black holes above the Planck mass \cite{FS97} . It is thus 
of interest to calculate more carefully the scale of ``passive'' fluctuations
due to stress tensor fluctuations.

\vskip .8cm
{\bf Acknowledgement:} This work was
supported in part by the National Science Foundation (Grant No. PHY-9800965).

\appendix
\section*{Appendix A}
\setcounter{equation}{0}
\renewcommand{\theequation}{A\arabic{equation}}

Assume there is a quantum state \( |\psi \rangle _{a} \) which can 
decompose
the field operator into positive and negative frequency parts \( \phi =\phi 
^{+}+\phi ^{-} \),
with \( \phi ^{+}|\psi \rangle_a =0 \).
By using Wick's theorem, the four-point function can be expressed as 
\begin{eqnarray}
\phi_1\phi_2\phi_3\phi_4 &=& N_a(\phi_1\phi_2\phi_3\phi_4)+ 
N_a(\phi_1\phi_2)\langle\phi_3\phi_4\rangle_a + 
N_a(\phi_1\phi_3)\langle\phi_2\phi_4 
\rangle_a+N_a(\phi_1\phi_4)\langle\phi_2\phi_3\rangle_a  \nonumber \\
&+& N_a(\phi_2\phi_3) 
\langle\phi_1\phi_4\rangle_a+N_a(\phi_2\phi_4)\langle\phi_1\phi_3\rangle_a 
+N_a(\phi_3\phi_4)\langle\phi_1\phi_2\rangle_a+\langle\phi_1\phi_2\rangle_a 
\langle\phi_3\phi_4\rangle_a \nonumber \\
&+& \langle\phi_1\phi_3\rangle_a\langle\phi_2\phi_4\rangle_a 
+\langle\phi_1\phi_4\rangle_a\langle\phi_2\phi_3\rangle_a \,,
\end{eqnarray}
where $N_a$ means normal ordering with respect to the state  
\( |\psi \rangle _{a} \)
and \( \langle \rangle _{a} \) means the expectation value in this  state.
If we take the expectation value of the above relation in the state \( |\psi \rangle _{a} \),
the result is
\begin{equation}
 \langle \phi_1\phi_2\phi_3\phi_4 
~\rangle_{a}=\langle\phi_1\phi_2\rangle_a\langle 
\phi_3\phi_4\rangle_a+\langle\phi_1\phi_3\rangle_a\langle\phi_2\phi_4 
\rangle_a+\langle\phi_1\phi_4\rangle_a\langle\phi_2\phi_3\rangle_a .
\end{equation} 
For the particular case of the Minkowski vacuum,
\begin{eqnarray}
\phi_1\phi_2\phi_3\phi_4 &=& :\phi_1\phi_2\phi_3\phi_4:+:\phi_1\phi_2: 
\langle\phi_3\phi_4\rangle_M+
:\phi_1\phi_3:\langle\phi_2\phi_4\rangle_M \nonumber \\
&+&:\phi_1\phi_4:\langle\phi_2\phi_3\rangle_M+
:\phi_2\phi_3:\langle\phi_1\phi_4\rangle_M 
+:\phi_2\phi_4:\langle\phi_1\phi_3\rangle_M \nonumber \\
&+& :\phi_3\phi_4:\langle\phi_1\phi_2\rangle_M 
+\langle\phi_1\phi_2\rangle_M\langle\phi_3\phi_4\rangle_M \nonumber \\
&+& \langle\phi_1\phi_3\rangle_M 
\langle\phi_2\phi_4\rangle_M+
\langle\phi_1\phi_4\rangle_M\langle\phi_2\phi_3\rangle_M \, ,
\end{eqnarray}
where $: :$ means normal order to the Minkowski vacuum state. and \( 
\langle \rangle _{M} \)
means the expectation value in the Minkowski vacuum.
The expectation value of the above equation in this state is
\begin{equation}
 \langle \phi_1\phi_2\phi_3\phi_4 
\rangle_{M}=\langle\phi_1\phi_2\rangle_{M}\langle\phi_3\phi_4\rangle_{M} 
+\langle\phi_1\phi_3\rangle_{M}\langle\phi_2\phi_4\rangle_{M}+ 
\langle\phi_1\phi_4\rangle_{M}\langle\phi_2\phi_3\rangle_{M} \, .
\end{equation} 
By using the expressions  
\begin{equation}
\phi_1\phi_2=:\phi_1\phi_2:+\langle\phi_1\phi_2\rangle_M
\end{equation}
and 
\begin{equation}
\langle\phi_1\phi_2\rangle_a=\langle:\phi_1\phi_2:\rangle_a+ 
\langle\phi_1\phi_2\rangle_M \, ,
\end{equation}
we  get
\begin{equation} 
\langle:\phi_1\phi_2:\rangle_a=\langle\phi_1\phi_2\rangle_a-
\langle\phi_1\phi_2\rangle_M 
\end{equation} 
and
\begin{equation} 
\langle:\phi_1\phi_2\phi_3\phi_4:\rangle_a=
\langle:\phi_1\phi_2:\rangle_a\langle 
:\phi_3\phi_4:\rangle_a+\langle:\phi_1\phi_3:\rangle_a\langle:\phi_2\phi_4: 
\rangle_a+\langle:\phi_1\phi_4:\rangle_a\langle:\phi_2\phi_3:\rangle_a \,.
\end{equation}

\vspace{0.2 cm}

\appendix
\section*{Appendix B}
\setcounter{equation}{0}
\renewcommand{\theequation}{B\arabic{equation}}

\subsection*{B1 Evaluation of \protect\( \int 
f(t)f(t')\frac{1}{(t-t')^{2}}dtdt'\protect \)}

The sampling function and its derivative are respectively 
\begin{equation}
f(t)=\frac{\tau}{\pi }\frac{1}{(t-t_{0})^{2}+\tau^{2}}
\end{equation}
and 
\begin{equation}
f'(t)=\frac{-2\tau}{\pi }\frac{t-t_{0}}{((t-t_{0})^{2}+\tau^{2})^{2}}.
\end{equation}
The integral after intergation by parts yields

\begin{eqnarray}
\int f(t)f(t')\frac{1}{(t-t')^{2}}dtdt' & = & \int \int 
f'(t)f'(t')\ln(t-t')dtdt'\nonumber \\
 & = & \int f'(t')dt'\int ^{\infty }_{-\infty 
}f'(t)\ln(t-t')dt\nonumber \\
 & = & \int f'(t')A(t')dt', 
\end{eqnarray}
where 
\begin{equation}
A(t')\equiv \frac{-2\tau}{\pi }\int ^{\infty }_{-\infty 
}\frac{t-t_{0}}{((t-t_{0})^{2}+\tau^{2})^{2}}\ln(t-t')dt.
\end{equation}
This integral contains a second order pole and can be done by residues.
 Let \( x=t-t_{0} \), and the integral becomes 
\begin{eqnarray}
A(t') & = & \frac{-2\tau}{\pi }\int ^{\infty }_{-\infty 
}\frac{x}{(x^{2}+\tau^{2})^{2}}\ln(x+t_{0}-t')dx\nonumber \\
 & = & \frac{-2\tau}{\pi }\oint 
\frac{z}{(z^{2}+\tau^{2})^{2}}\ln(z+t_{0}-t')dz\nonumber \\
 & = & \frac{-2\tau}{\pi }\left(-\frac{\partial }{\partial \tau^{2}}\right)
\oint \frac{z}{z^{2}+\tau^{2}}\ln(z+t_{0}-t')\nonumber \\
 & = & \frac{2\tau}{\pi }(2\pi i)\left(\frac{\partial }{\partial \tau^{2}}\right)
 \left[\frac{\ln (z+t_{0}-t')}{z+i \tau}\right]_{z=i\tau }\nonumber \\
 & = & \frac{b}{\tau^{2}+b^{2}}+i\frac{\tau}{\tau^{2}+b^{2}},
\end{eqnarray}
 where \( R=\sqrt{\tau^{2}+b^{2}} \) and \( b=t'-t_{0} \) . The double 
integral becomes
\begin{equation}
\int f'(t')A(t')dt'=\frac{-2\tau}{\pi }\frac{\pi 
[4\tau^{2}-(t_{0}-t_{0}')^{2}]}{2\tau[4\tau^{2}+(t_{0}-t_{0}')^{2}]^{2}} + 
i\left[\int f'(t')\, Im[A(t')] dt'\right]
\end{equation}
We keep only the real part and write
\begin{equation}
\int f(t)f(t')\frac{1}{(t-t')^{2}}dtdt'=\frac{-2\tau}{\pi }\frac{\pi 
[4\tau^{2}-(t_{0}-t_{0}')^{2}]}{2\tau [4\tau^{2}+(t_{0}-t_{0}')^{2}]^{2}}.
\end{equation}
In the coincidence limit \( t_{0}\to t_{0}' \), the integral is 
\begin{equation}
\lim _{t_{0}'\to t_{0}}\int 
f(t)f(t')\frac{1}{(t-t')^{2}}dtdt'=-\frac{1}{4\tau^{2}}. \label{eq:App_B1}
\end{equation}

\subsection*{B2 Evaluation of \protect\( \int 
f(t)f(t')\frac{1}{(t-t')^{4}}dtdt'\protect \) }

The second derivative of sampling function is 
\begin{equation}
f''(t)=-\frac{2\tau}{\pi} 
\left[\frac{1}{((t-t_{0})^{2}+\tau^{2})^{2}}-
\frac{4(t-t_{0})^{2}}{((t-t_{0})^{2}+\tau^{2})^{3}}\right] \, .
\end{equation}
Integrating by parts yields
\begin{equation}
\int f(t)f(t')\frac{1}{(t-t')^{4}}dtdt'=\frac{-1}{6}\int \int 
f(t)''f(t')'' \ln(t-t')dtdt'.
\end{equation}
A similar calculation as in B1 yields
\begin{eqnarray}
B(t') & = & \int ^{\infty }_{-\infty }f''(t)\ln(t-t')dt\nonumber \\
 & = & \frac{2\tau}{\pi }\left[\frac{\partial }{\partial \tau^{2}}\oint 
\frac{\ln z}{(z-t'-t_{0})^{2}+\tau^{2}}dz+2\left(\frac{\partial }{\partial 
\tau^{2}}\right)^{2}\oint 
\frac{(z+t'-t_{0})^{2}}{(z+t'-t_{0})^{2}+\tau^{2}}\ln zdz\right]\nonumber \\
 & = & \frac{2a}{\pi }\left[\frac{\partial }{\partial 
\tau^{2}}B_{1}+
2\left(\frac{\partial }{\partial \tau^{2}}\right)^{2}B_{2}\right].
\end{eqnarray}
where
\begin{equation}
B_{1}=\frac{\pi }{\tau}\ln z_{0}
\end{equation}
and 
\begin{equation}
B_{2}=-\pi \tau \ln z_{0},
\end{equation}
where \( z_{0}=\sqrt{\tau^{2}+b^{2}} \) and \( b=t'-t_{0} \). The 
integral \( B(t') \) becomes
\begin{equation}
B(t')=\int ^{\infty }_{-\infty 
}f''(t)\ln(t-t')dt=-\frac{1}{\tau^{2}+b^{2}}+
\frac{2\tau^{2}}{(\tau^{2}+b^{2})^{2}} \, .
\end{equation}
Substituting this result into the original double integral yields

\begin{eqnarray}
 &  & \int f(t)f(t')\frac{1}{(t-t')^{4}}dtdt'\nonumber \\
 & = & \frac{-\tau}{3\pi }\int ^{\infty }_{-\infty 
}(\frac{1}{((t'-t_{0}')^{2}+\tau^{2})^{2}}-
\frac{4(t'-t_{0}')^{2}}{((t'-t_{0}')^{2}
+\tau^{2})^{3}})(\frac{1}{\tau^{2}+b^{2}}-
\frac{2\tau^{2}}{(\tau^{2}+b^{2})^{2}})dt'  \nonumber 
\\
 & = & 
\frac{16\tau^{2}-24\tau^{2}(t_{0}-t_{0}')^{2}+(t_{0}-t_{0}')^{4}}{(4\tau^{2}+
(t_{0}-t_{0}')^{2})^{4}}.
\end{eqnarray}
In the limit \( t_{0}-t_{0}' \rightarrow0 \),  
the integral becomes 
\begin{equation}
\lim _{t_{0}'\to t_{0}}\int 
f(t)f(t')\frac{1}{(t-t')^{4}}dtdt'=\frac{1}{16\tau^{4}}. \label{eq:App_B2}
\end{equation}

\end{document}